\UseRawInputEncoding
\documentclass[aps,prb,twocolumn,showpacs,psfig,superscriptaddress,longbibliography]{revtex4-1}

\pdfoutput=1
\usepackage{textcomp}
\usepackage{times}
\usepackage{graphicx}
\usepackage{float}
\usepackage{latexsym,amsmath,amssymb,bm,euscript}
\usepackage{color}
\usepackage{subfigure}
\usepackage{epstopdf}
\usepackage[colorlinks=true,linkcolor=blue,citecolor=blue]{hyperref}
\usepackage{hyperref}
\usepackage{soul}
\usepackage[normalem]{ulem}
\usepackage{mathrsfs}
\usepackage{amsmath}
\usepackage{lettrine}
\usepackage{xspace}
\usepackage{textcomp}

\begin{document}

\hyphenpenalty=5000
\tolerance=1000

\title{Low-energy spin dynamics in a Kitaev material Na$_3$Ni$_2$BiO$_6$ investigated by NMR}	

\author{Xinyu Shi}
\thanks{These authors contributed equally to this study.}
\affiliation{Mathematics and Physics Department,~North China Electric Power University,~Beijing,~102206,~China}

\author{Yi Cui}
\thanks{These authors contributed equally to this study.}
\affiliation{Department of Physics and Beijing Key Laboratory of
	Opto-electronic Functional Materials $\&$ Micro-nano Devices,~Renmin
	University of China,~Beijing,~100872,~China}
\affiliation{Key Laboratory of Quantum State Construction and
Manipulation (Ministry of Education),~Renmin University of China, Beijing, 100872,~China}

\author{Yanyan Shangguan}
\thanks{These authors contributed equally to this study.}
\affiliation{National Laboratory of Solid State Microstructures and Department of Physics,~Nanjing University,~Nanjing, 210093,~China}

\author{Xiaoyu Xu}
\affiliation{Department of Physics and Beijing Key Laboratory of
	Opto-electronic Functional Materials $\&$ Micro-nano Devices,~Renmin
	University of China,~Beijing,~100872,~China}

\author{Zhanlong Wu}
\affiliation{Department of Physics and Beijing Key Laboratory of
	Opto-electronic Functional Materials $\&$ Micro-nano Devices,~Renmin
	University of China,~Beijing,~100872,~China}

\author{Ze Hu}
\affiliation{Department of Physics and Beijing Key Laboratory of
	Opto-electronic Functional Materials $\&$ Micro-nano Devices,~Renmin
	University of China,~Beijing,~100872,~China}

\author{Shuo Li}
\affiliation{Department of Physics and Beijing Key Laboratory of
	Opto-electronic Functional Materials $\&$ Micro-nano Devices,~Renmin
	University of China,~Beijing,~100872,~China}

\author{Kefan Du}
\affiliation{Department of Physics and Beijing Key Laboratory of
	Opto-electronic Functional Materials $\&$ Micro-nano Devices,~Renmin
	University of China,~Beijing,~100872,~China}

\author{Ying Chen}
\affiliation{Department of Physics and Beijing Key Laboratory of
	Opto-electronic Functional Materials $\&$ Micro-nano Devices,~Renmin
	University of China,~Beijing,~100872,~China}

\author{Long Ma}
\affiliation {Anhui Province Key Laboratory of Condensed Matter Physics at Extreme Conditions,~High Magnetic Field Laboratory,~Chinese Academy of Sciences,~Hefei,~230031,~China}

\author{Zhengxin Liu}
\affiliation{Department of Physics and Beijing Key Laboratory of
	Opto-electronic Functional Materials $\&$ Micro-nano Devices,~Renmin
	University of China,~Beijing,~100872,~China}
\affiliation{Key Laboratory of Quantum State Construction and
Manipulation (Ministry of Education),~Renmin University of China, Beijing, 100872,~China}

\author{Jinsheng Wen}
\email{jwen@nju.edu.cn}
\affiliation{National Laboratory of Solid State Microstructures and Department of Physics,~Nanjing University,~Nanjing, 210093,~China}
\affiliation{Innovative Center for Advanced Microstructures, Nanjing University, Nanjing, 210093, China}

\author{Jinshan Zhang}
\email{zhangjs@ncepu.edu.cn}
\affiliation{Mathematics and Physics Department,~North China Electric Power University,~Beijing,~102206,~China}

\author{Weiqiang Yu}
\email{wqyu\_phy@ruc.edu.cn}
\affiliation{Department of Physics and Beijing Key Laboratory of
	Opto-electronic Functional Materials $\&$ Micro-nano Devices,~Renmin
	University of China,~Beijing,~100872,~China}
\affiliation{Key Laboratory of Quantum State Construction and
Manipulation (Ministry of Education),~Renmin University of China, Beijing, 100872,~China}

\begin{abstract}
We performed $^{23}$Na NMR and magnetization measurements on an $S$~=~1,  quasi-2D honeycomb lattice antiferromagnet Na$_3$Ni$_2$BiO$_6$.
A large positive Curie-Weiss constant of 22.9~K  is observed.
The NMR spectra at low fields are consistent with a ``zigzag" magnetic order,
indicating a large easy-axis anisotropy.
With field applied along the $c^{*}$ axis, the NMR spectra
confirm the existence of a 1/3-magnetization plateau phase between 5.1~T and 7.1~T.
The transition from the zigzag order to the 1/3-magnetization plateau phase is also found to be a first-order type.
A monotonic decrease of the spin gap is revealed in the 1/3-magnetization plateau phase,
which reaches zero at a quantum critical field  $H_{\rm C}$$\approx$~8.35~T before entering
the fully polarized phase. These data suggest the existence of exchange frustration
in the system along with strong ferromagnetic interactions, hosting the possibility for Kitaev physics.
Besides, well below the ordered phase, the 1/$T_1$ at high fields shows either a
level off or an enhancement upon cooling below 3~K, which suggests the existence of low-energy fluctuations.


\end{abstract}

\maketitle

\section{Introduction}


The spin-1/2 Kitaev model on a honeycomb lattice is a rare case of an exactly solvable model for a quantum spin liquid (QSL) with Dirac Majorana fermions and gauge fields, and provides a unique pathway
for quantum computations~\cite{Kitaev_AP_2006}, which has attracted significant research attention recently.
In real materials, Kitaev interactions, namely the $K$ term, are believed to
arise from frustrated exchange couplings through strong spin-orbit coupling and bond symmetry, allowing materials to
potentially exhibit Kitaev QSL~\cite{Khaliullin_PRL_2009,Kee_PRL_2014,Kindo_PRB_2016}.
The so-called Kitaev materials, notable Na$_2$IrO$_3$~\cite{Hill_PRB_2011,Cao_PRB_2012,Taylor_PRL_2012}
and ${\alpha}$-RuCl$_3$~\cite{Coldea_PRB_2015,Kim_PRB_2015} with layered honeycomb lattices,
are effective spin-1/2 systems consisting of Kramers doublet from a d$^{5}$ electronic configuration
with local edge-shared octahedral structure and strong spin-orbit coupling.
Later studies on cobaltates, also with the
layered honeycomb lattice, revealed that when the non-octahedral crystal field is sufficiently weak to preserve
spin-orbit entanglement, Co$^{2+}$ also possesses a pseudo-spin 1/2 degree of freedom~\cite{Simonet_PRB_2016,Ritter_PRB_2017,Ling_JSSC_2016,McGuire_PRM_2019,Park_JPCM_2022}
and potentially provides Kitaev interactions~\cite{Khaliullin_PRL_2009,Kee_PRL_2014,Park_JPCM_2022}.

In all above compounds, the low-temperature ground state, however, is not a QSL, but rather exhibits a long-range magnetic order, commonly
identified as a ``zigzag" antiferromagnetic (AFM) order. Interestingly, later high-field studies on
these materials revealed that magnetic field, by suppressing magnetic ordering, may help to induce putative QSL phases,
such as that observed in ${\alpha}$-RuCl$_3$~\cite{JiachengZheng_PRL_2017,Nagler_NM_2016,lee_PRL_2017,Kim_PRB_2017,Loidl_PRL_2017,Nagler_PRB_2019,Ong_NP_2021,Li_NC_2021},
Na$_2$Co$_2$TeO$_6$~\cite{Li_PRB_2020,Ma_NC_2021} and Na$_3$Co$_2$SbO$_6$~\cite{Li_PRX_2022,ZeHu_PRB_2024}.

Recently, magnetization measurements on  Na$_3$Ni$_2$BiO$_6$, a spin-1 Ising antiferromagnet with layered honeycomb lattice,
demonstrates a one-third magnetization plateau (1/3-MP) phase with an out-of-plane field, which clearly indicates the existence of magnetic frustration~\cite{Wen_NP_2023}.
The fractional MP is an interesting phenomenon that usually manifests in systems with magnetic frustration, which is further enhanced by quantum effect in small spin systems, such as
spin-1/2 antiferromagnets with triangular~\cite{Winkelmann_JPCM_1994,Goto_PRB_2003,Takano_PRB_2007,Takano_PRL_2009,Kindo_PRL_2012,Takano_PRL_2012,Tanaka_PRL_2013,Ma_NC_2018}
, kagome~\cite{Zhitomirsky_PRL_2002,Senthil_PRL_2006,Hotta_NC_2013,Kumar_PRB_2021,Suwa_PRB_2022}, and $J_1$-$J_2$ square lattices~\cite{Notych_SSC_1993,Ueda_PRL_1999,Mila_Sci_2002},
where the MP phases can be explained by the ``order by disorder'' mechanism~\cite{Golosov_JPCM_1991,Honecker_JPCM_1999,Starykh_RPP_2015,Petrenko_PRL_2000,Miyashita_JPSJ_1985,Henley_PRL_1989,Starykh_PRL_2009,Mila_PRB_2013,Danshita_PRL_2014}.
The discussions have been extended to some triangular-lattice systems with higher spins~\cite{Goto_JPSJ_1996,Nakano_JPSJ_2011,Schlottmann_PRL_2012}, and
for classical spins, the 1/3-MP phase is proposed to survive in an enlarged field range as temperature rises above zero~\cite{Shannon_PRB_2011}.
Attempts have also been made in honeycomb-like antiferromagnets~\cite{Kindo_PRB_2016,Hagiwara_JPSJ_2019}.
Elastic neutron scattering studies on Na$_3$Ni$_2$BiO$_6$ revealed a zigzag order with local moments along the crystalline $c^{*}$ axis at zero field, and a field-induced
partial spin-flop phase, which indicates the existence of magnetic frustration~\cite{Wen_NP_2023}.
Density functional theory (DFT) calculations on Na$_3$Ni$_2$BiO$_6$ suggest that anisotropic
Kitaev interactions may exist and play a key role in the formation of the MP phase.
This extends the study of exchange frustration, particularly Kitaev interactions, to high spin systems~\cite{Wen_NP_2023}.

To reveal the spin dynamics in this system, we performed $^{23}$Na NMR measurements on Na$_3$Ni$_2$BiO$_6$.
Along with magnetization measurements, our main results are summarized in the phase diagram shown in Fig.~\ref{pd}.
The 1/3-MP phase is reconstructed by the shift of the NMR resonance frequency.
By measuring the spin-lattice relaxation rates $1/^{23}T_1$, all the phase boundaries between long-range ordered state and paramagnetic (PM) state
in the whole field range are determined by extracting the N\'{e}el temperatures $T_N$.
Furthermore, a spin gap is found, which decreases monotonically with field and diminishes at a critical field of 8.35~T.
With further increase of field, the gap increases towards the fully polarized phase.
Surprisingly, with phases deep in the ordered phase and the fully polarized phase, a gapless
behavior is seen in the low-temperature $1/^{23}T_1$, which suggests the existence of strong low-energy dynamics.
We speculate that structural dynamics caused by Na movement, rather than intrinsic spin dynamics, still persist at such low temperatures in the layered compound.

\section{Materials and Techniques}

\begin{figure}[t]
\includegraphics[width=8.5cm]{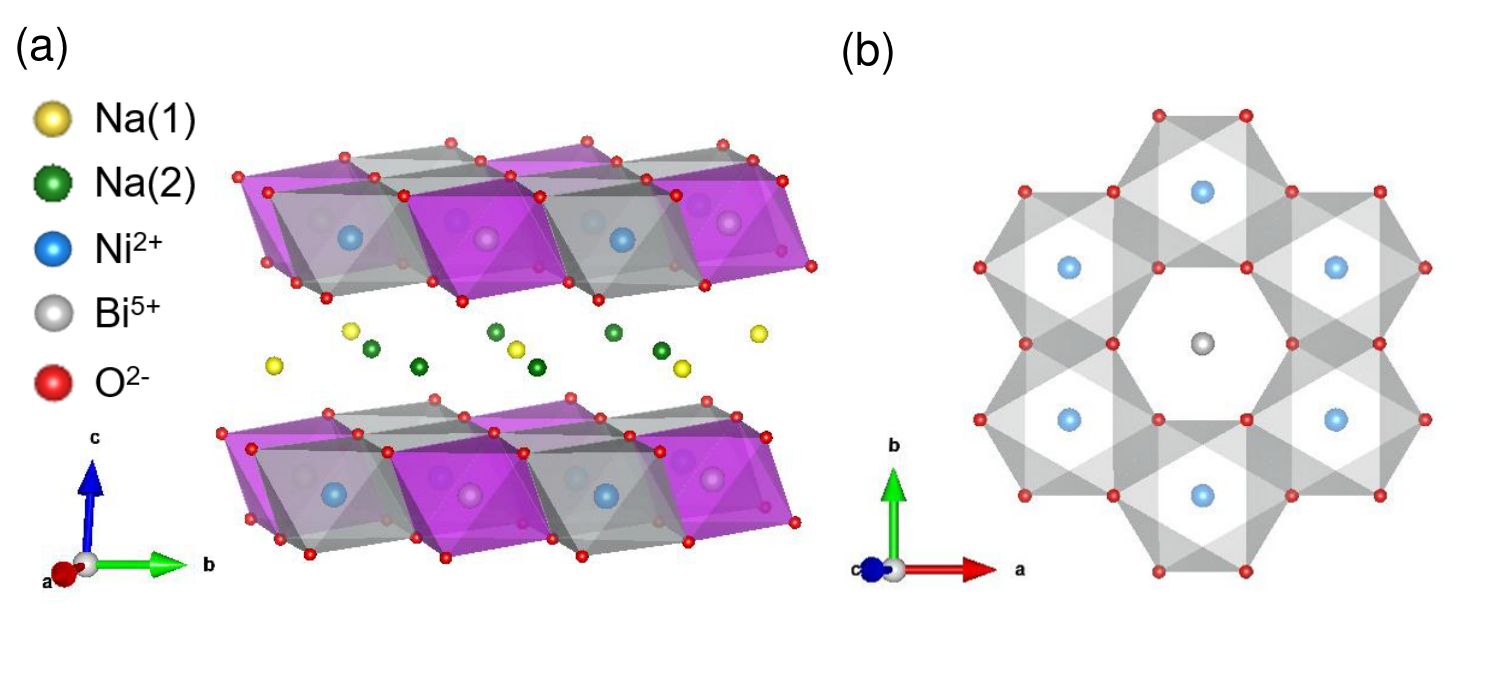}
\caption{\label{struc}~\textbf{Crystal of Na$_3$Ni$_2$BiO$_6$.}
(a) A schematic drawing of the 3D crystal structure.
(b) Top view of the NiO$_6$ octahedra forming the honeycomb layers.
}
\end{figure}

The Na$_3$Ni$_2$BiO$_6$ compound crystallizes in a monoclinic structure with a layered arrangement, belonging to the space group $C_2$/m~\cite{Cava_Inorg_2013}, as shown in Fig.~\ref{struc}(a). NiO$_6$ and BiO$_6$
octahedra are arranged in a 2:1 ratio in the $ab$ plane, with sodium ions located above the octahedra among the layers.
Ni$^{2+}$ ions form the spin-1, quasi-two-dimensional structure with a layered honeycomb lattice. Two types of
inequivalent Na sites, Na(1) and Na(2) with a 1:2 occupancy ratio, reside between the honeycomb layers.

High quality single crystals were grown by the flux method~\cite{Cava_Inorg_2013}, with typical sizes of
2mm*4mm*0.5mm.  Bulk susceptibility and magnetization
measurements were performed in a Quantum Design Physical Property Measurement System (PPMS) with temperatures down to 2~K and fields up to 14~T applied perpendicular to the $ab$ plane, that is, along the crystalline $c^{*}$ axis.
We conducted NMR measurements on $^{23}$Na nuclei, which possess spin $I$ = 3/2 and Zeeman factor $\gamma$$=$~11.262 MHz/T.
The field was also applied perpendicular to the $ab$ plane, with field up to 12~T and temperature down to 1.5~K.
Note that magnetic hysteresis was seen at intermediate temperatures, so most magnetization and NMR measurements were performed with increasing field. The magnetic field was calibrated by the resonance frequency of $^{63}$Cu from the NMR coil.

The spectra were obtained by the standard spin-echo technique, with typical $\pi/2$ time length of 2$\mu$s. For
broad spectra at low temperatures, the whole spectra were acquired by sweeping frequencies.
The NMR Knight shift was calculated by $K_n$$=$$(f/{\gamma}H-1){\times}100\%$, where $f$ is the frequency of the center peak.
The spin-lattice relaxation time $^{23}T_1$ was measured by the spin inversion-recovery method, with the nuclear magnetization
fit to the exponential function, $I(t)$$=~$$a-b[e^{-(t/T_1)^\beta}+9e^{-(6t/T_1)^\beta}]$, where $\beta$
is a stretching factor. In the PM phase, $\beta$$\approx$~1, indicating high quality of the crystals, and decreases in the ordered phase. Note that with $1/T_1$$=~$$T\lim_{\omega{\to}0}\sum_q~A^2_{hf}(q){\text{Im}~\chi(q,\omega)}/{\omega}$, $1/T_1$ measures the low-energy spin dynamics, where $\chi(q,\omega)$ represents the dynamical susceptibility, and $A_{hf}(q)$ is the hyperfine coupling constant.

\section{Susceptibility and Magnetization data}

\begin{figure}[t]
 \includegraphics[width=8.5cm]{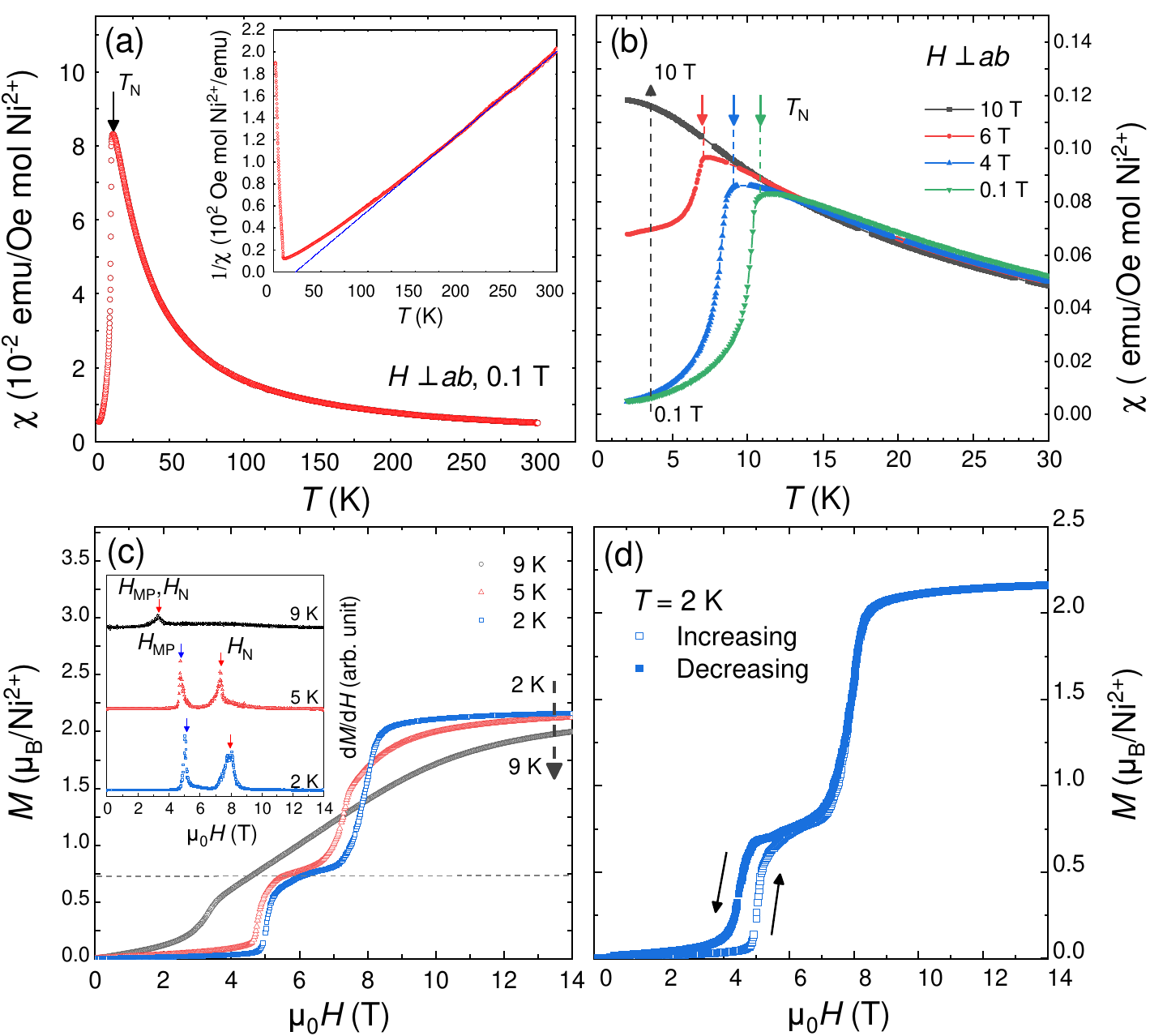}
   \caption{\label{sus}~\textbf{Magnetic susceptibility and magnetization.} (a) Magnetic susceptibility $\chi$ with an out-of-plane field of 0.1~T.
The black arrow marks the peak feature at $T_N$. Inset:  $1/\chi$ as a function of temperature. The straight line represents a linear fit
to $1/\chi$ with temperature from 150~K to 300~K.
(b) Magnetic susceptibility as a function of temperature at different fields. The arrows mark the $T_N$ at the peak positions of $\chi$.
(c) Magnetization as functions of fields, measured at different temperatures. The dotted horizontal line is placed at 1/3 of the saturated magnetic moment.
Inset: $dM/dH$ at different temperatures, with arrows placed at peaked positions, marking transitions to the 1/3-MP phase at $H_{\rm MP}$ and to the paramagnetic phase at $H_{\rm N}$, respectively.
(d) The magnetic hysteresis loop of $M(H)$ measured at 2~K.
}
\end{figure}

We initially examined the crystal through susceptibility and magnetization measurements with out-of-plane fields, specifically along the magnetic easy axis~\cite{Wen_NP_2023}.
Figure~\ref{sus}(a) shows the DC susceptibility $\chi$ as a function of temperature, measured at 0.1~T.
$\chi$ first increases upon cooling from 300~K to 10 K. A peaked behavior is seen at 11.6$\pm$0.1~K,
marking $T_N$  of the N\'{e}el ordering. Below $T_N$, $\chi$ drops rapidly in the ordered phase.

The illustration in Fig.~\ref{sus}(a) displays the inverse susceptibility $1/\chi$ at this low field,
which follows a linear temperature dependence at high temperatures, as expected for
a Curie-Weiss behavior. A linear fit, as shown by the solid straight line,
gives a positive  Curie-Weiss temperature $\theta_{CW}$$\approx$~22.9$\pm$0.1~K, larger than the previously reported value
of 13~K~\cite{Cava_Inorg_2013}. The large positive hyperfine coupling indicates strong ferromagnetic exchange coupling.

In Fig.~\ref{sus}(b), we further present $\chi$ at different fields as a function of temperature.
As the magnetic field increases, $T_N$, marked at the peak position of $\chi$, shift towards
lower temperatures. At 10~T, $\chi$  levels off at low temperatures without any peaked
behavior, indicating the onset of fully polarized phase.
The $T_N$ at different fields are then determined and plotted in Fig.~\ref{pd}.

At fields from 0.1~T to 4~T, $\chi$ drops to nearly zero at 2~K, consistent
with the zigzag order with moments perpendicular to the $ab$ plane.
At 6~T, however, $\chi$ first drops below $T_N$, and then approaches a large value at 2~K,
which indicates a different magnetic ground state.
The magnetization $M(H)$ was then measured at different temperatures with increasing fields,
following the earlier study~\cite{Wen_NP_2023}.
Figure~\ref{sus}(c) displays the $M(H)$ as functions of fields at $T=$~2~K, 5~K, and 9~K.
At intermediate fields, a magnetization plateau is clearly seen,
with the value close to one-third of the saturated value (2.13~$\mu_B$) of
Ni$^{2+}$. For guidance, a horizontal one-third magnetization line
is placed at 0.73~${\mu}_B$, which clearly indicates the onset of a
1/3-MP phase with fields from 5.1~T to 7.1~T.

In order to precisely identify the transition field, the differential of magnetization
d$M$/d$H$ at different temperatures are calculated and plotted in the inset of Fig.~\ref{sus}(c).
Distinctly, at both 2~K and 5~K, there are two peaks representing the upper and lower boundary fields of the MP phase.
The lower one, labeled as $H_{\rm MP}$, marks the transition from the zigzag order
to the MP phase. With increasing temperature, $H_{\rm MP}$ barely shifts,
which is typical for the MP phase.
The upper one, as labeled by $H_{\rm N}$, is the transition from the
magnetically ordered phase to the PM phase. By contrast, $H_{\rm N}$ shifts
largely with temperature, which corresponds to the N\'{e}el transition.
Both $H_{\rm MP}$ and $H_{\rm N}$, along with their corresponding temperatures, are
then added in the ($H$, $T$) phase diagram shown in Fig.~\ref{pd}.
The field regime of 1/3-MP phase shrinks with increasing temperature,
indicating a strong effect of quantum fluctuations rather than thermal fluctuations~\cite{Cava_Inorg_2013}.

The $M(H)$ also shows an obvious hysteresis loop at 2~K, through the transition
from the zigzag order to the 1/3-MP phase, as shown in Fig.~\ref{sus}(d). The presence of a hysteresis loop is a clear signature of a
first-order phase transition, which can occur between two phases both with Ising-type symmetry breaking,
that is, the moments are ordered along the $c^*$ axis as suggested by the neutron scattering data~\cite{Wen_NP_2023}. Our $M(H)$ data with
rising field is consistent with the previous report~\cite{Wen_NP_2023}, although the later,
lacking hysteresis data, concluded a second-order phase transition.

\section{Low-field NMR spectra and Knight shift}

\begin{figure}[t]
\includegraphics[width=8.5cm]{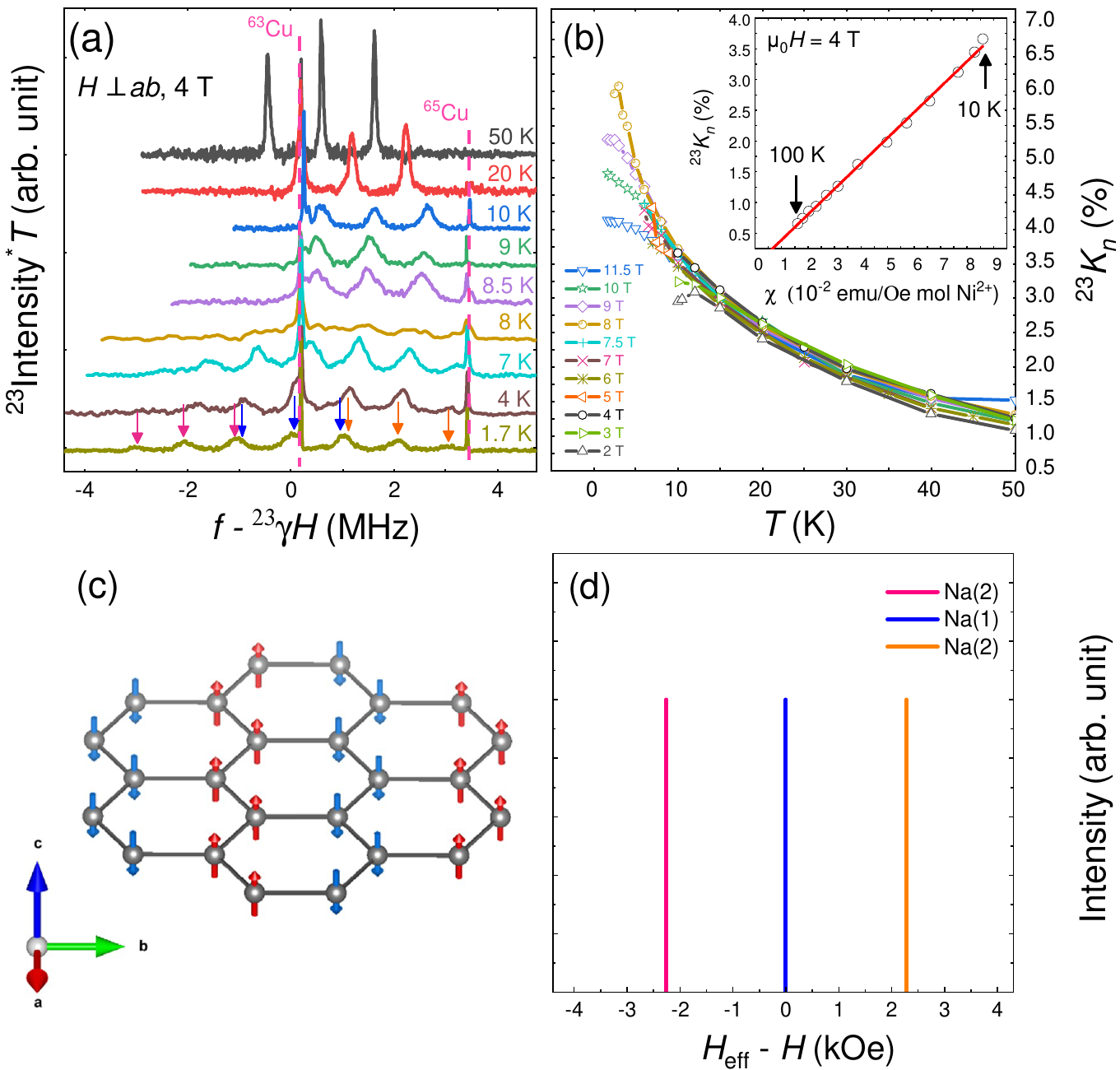}
\caption{\label{spec}~\textbf{Low-field $^{23}$Na NMR spectra with perpendicular field.}
(a) The $^{23}$Na NMR spectra at a fixed field of 4~T at different temperatures.
Vertical offsets are applied for clarity. Each spectrum above $T_N$ consists one center and two satellite peaks.
Data at 1.7~K are attributed to three sets of spectra, with red, blue and orange corresponding to three types of sites with different hyperfine field,
and each contributing one center peak and two satellite peaks. The sharp peaks are $^{63}$Cu and $^{65}$Cu resonance lines from the NMR coil.
(b) The Knight shift $^{23}K_n$ as functions of temperatures at typical fields. Inset: $^{23}K_n$ as a function of $\chi$ at a field of 4~T, with a
straight line presenting a linear fit.
(c) Illustration of zigzag magnetic structure, with moments along the $c^*$ axis.
(d) Simulated center NMR peaks of $^{23}$Na in the zigzag state. Na(2) has line splits while Na(1) does not.
}
\end{figure}

The $^{23}$Na NMR spectra of Na$_3$Ni$_2$BiO$_6$ at a low field of 4~T are demonstrated in Fig.~\ref{spec}(a) at different temperatures.
In the high-temperature PM state above 9~K, three NMR peaks are seen in the spectra, which
are attributed to one central peak and two satellite peaks on two sides, with the $\nu_q$$\approx$~1~MHz. Note that Na(1)
and Na(2) cannot be separated in PM phase, indicating that the difference in the hyperfine field and EFG (the Electron Field Gradient) on both nuclear
sites is small with both types of nucleus residing among the honeycomb layers.
The Knight shift $^{23}K_n$ is calculated from the resonance frequency
of the center peaks. $^{23}K_n$ is then plotted as functions of temperatures at different fields, rising upon cooling and characterizing the PM behavior, as shown in Fig.~\ref{spec}(b).

The Knight shift $K_n$$=$$K_c+K_s$, which contains a constant chemical and a spin contribution as specified by the last two terms respectively.
The spin contribution, $K_s$($T$), scales linearly with bulk susceptibility, $\chi$($T$),
given by $K_s = {A_{hf}\chi}/{N_A\mu_B}$, where $N_A$ is the Avogadro constant, ${\mu_B}$ is the Bohr magneton, and $A_{hf}$ is the hyperfine coupling constant.
With this, we plot $K_n$ as a function of $\chi$, both measured at 4~T with temperature from 10~K to 100~K, where a linear fit is clearly
obeyed, as shown in the inset of Fig.~\ref{spec}(b). Note that as  $\chi(T)$ follows a Curie-Weiss fit in
a large temperature from 300~K down to 150~K, the $\chi(T)$ measured on the sample has no orbital contribution, or at least negligible. From the linear fit, we obtain a very small orbital contribution
$K_c$$\approx$~$-0.026\%$, and the hyperfine coupling constant $A_{hf}$$=~$2.42$\pm$0.03~kOe/$\mu_B$.

The spectrum at 8 K broadens significantly (Fig.~\ref{spec}(a)), which indicates the magnetic phase transition where the local field becomes inhomogenous across the sample. At 7~K, clear peaks are seen again, and when cooled down to 1.7~K, seven
NMR lines are resolved. The splits of these lines should indicate the onset of AFM ordering.
Note that with $\nu_q$$\approx$~1~MHz determined at hight temperature, we separate seven peaks
into three sets, as marked by red, blue and orange down arrows on the spectrum at 1.7~K. Each set contains
three peaks, with almost equal intensity and separated by about 1~MHz, as expected for one center peak and two satellite peaks.

In fact, such assignment of NMR peaks is consistent with the zigzag order. Figure~\ref{spec}(c) illustrates
the magnetic structure of Ni$^{2+}$ moments in the zigzag AFM phase as revealed by the neutron scattering data~\cite{Wen_NP_2023}.
With the above magnetic configuration and the local moment of 2$\mu_B$/Ni$^{2+}$, we simulated the center peaks of the $^{23}$Na spectra,
by considering the dipolar coupling between $^{23}$Na nuclei and Ni$^{2+}$  moments. The resulting $^{23}$Na resonance lines are plotted against
$H_{\rm eff}-H$, where $H_{\rm eff}$ is the total field seen by each nucleus, and $H$ is the static field applied perpendicular to the $ab$ plane, as shown in Fig.~\ref{spec}(d).
The simulation gives a single Na(1) line, and two split Na(2) lines due to zigzag order, with the ratio of the spectral weight 1:1:1.
The three-line feature is consistent with the measured spectra at 1.7~K, with three sets of spectra (each containing one center peak).
The simulation gives separated center lines with frequencies of -2.26~kOe (-2.54~MHz), 0~kOe (0~MHz) and 2.28~kOe (2.56~MHz),
which are close to the measured values.

\section{high-field NMR spectra and magnetization plateau phase}

\begin{figure}[t]
\includegraphics[width=8.5cm]{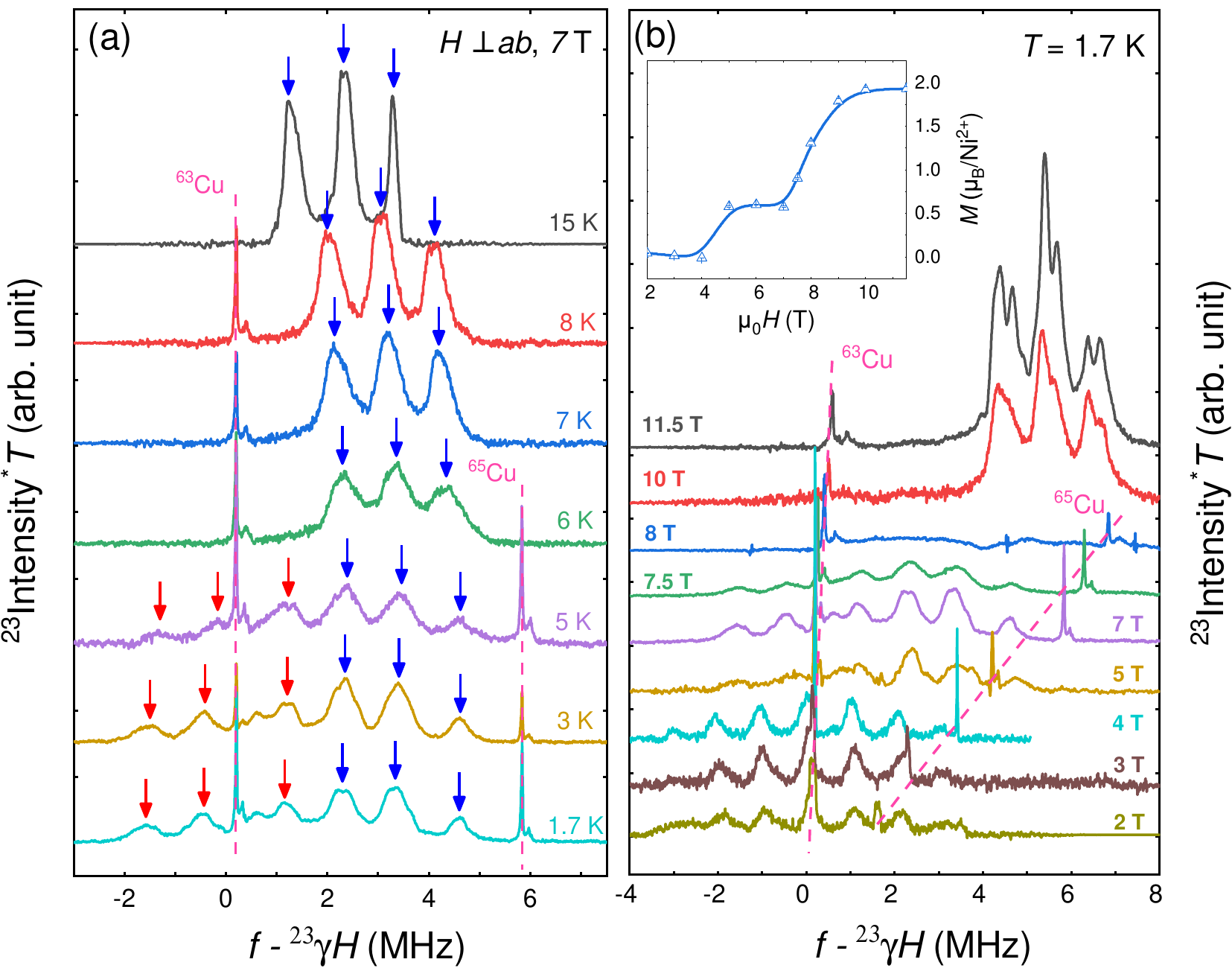}
\caption{\label{spec2}~\textbf{High-field $^{23}$Na spectra with perpendicular field.}
(a) Spectra at a fixed field of 7~T measured at typical temperatures. Blue and red arrows mark two sets
of peaks, each with one center line and two satellite lines.
(b) Spectra measured at 1.7~K with increasing field. Inset: Calculated longitudinal magnetization from the NMR spectra (see text).}
\end{figure}

Figure~\ref{spec2}(a) displays the spectra at a high field of 7~T, measured with decreasing temperature. The spectra above 5.5~K ($T_N$ at 7~T) characterized
by three peaks (marked by blue arrows) with one center and two satellite lines. Below $T_N$, a new set of lines with three
peaks (marked by red down arrows) emerge at lower frequencies with smaller spectral weight.
At 1.7~K, the center line of the low-frequency set reads a negative frequency about -0.5~MHz, and the
center line of high-frequency one reads a large positive frequency about 3.3~MHz, which indicates again the onset of an AFM ground state.
However this spectrum is distinctive to that observed at 4~T, 1.7~K (Fig.~\ref{spec}~(a)).

For comparison, the spectra at 1.7~K were measured and compared at different fields, as shown in Fig.~\ref{spec2}~(b).
Three types of ground states are clearly resolved  by the change of the NMR lineshape with increasing field.
Below 4~T, the $^{23}$Na spectra exhibit three sets of lines as discussed before; at 5 to 7~T, two sets of lines appear
with a large spectral weight at the high frequency side; above 9~T, three NMR lines emerge,
indicating the fully polarized phase characterized by one center line and two satellite lines with
a uniform hyperfine field on all Na sites. At 10~T and 11.5~T, a small line split is observed, which should be attributed to the
resolved Na(1) and Na(2) lines when the hyperfine field becomes very large in the polarized phase.

Indeed, the total longitudinal magnetization can also be obtained from the NMR spectra. By calculating the
average frequency of the $^{23}$Na spectra, with the value of $A_{hf}$ shown earlier,
$M$-$H$ is generated and shown in the inset of Fig.~\ref{spec2}(b).
Clearly, the 1/3-MP phase is established at field from 5~T - 7~T.

To further support the MP phase, we state that with a large positive resonance frequency observed at
7~T and 1.7~K (Fig.~\ref{spec2}~(a)), a strong local hyperfine field is suggested.
Indeed, this is consistent with the existence of clustered up moments ($\uparrow\uparrow$),
which should produce a large positive hyperfine field as observed.
Based on the neutron scattering data, or theoretical proposals, many magnetic structures for the 1/3 MP phase have been proposed
but have not yet been resolved, including  $\circ\uparrow\circ\downarrow\uparrow\uparrow$  and
$\downarrow\downarrow\uparrow\uparrow\uparrow\uparrow$, all having enlarged magnetic unit~\cite{Wen_NP_2023}.
Unfortunately, our simulation of the spectra based on any of these patterns (data not shown) failed to
produce the observed spectra, which may be caused by our oversimplified assumption
of pure dipolar hyperfine coupling among nuclear spins and the magnetic moments,
or unknown tilting of local moment toward the out-of-plane direction.

\begin{figure}[t]
\includegraphics[width=7cm]{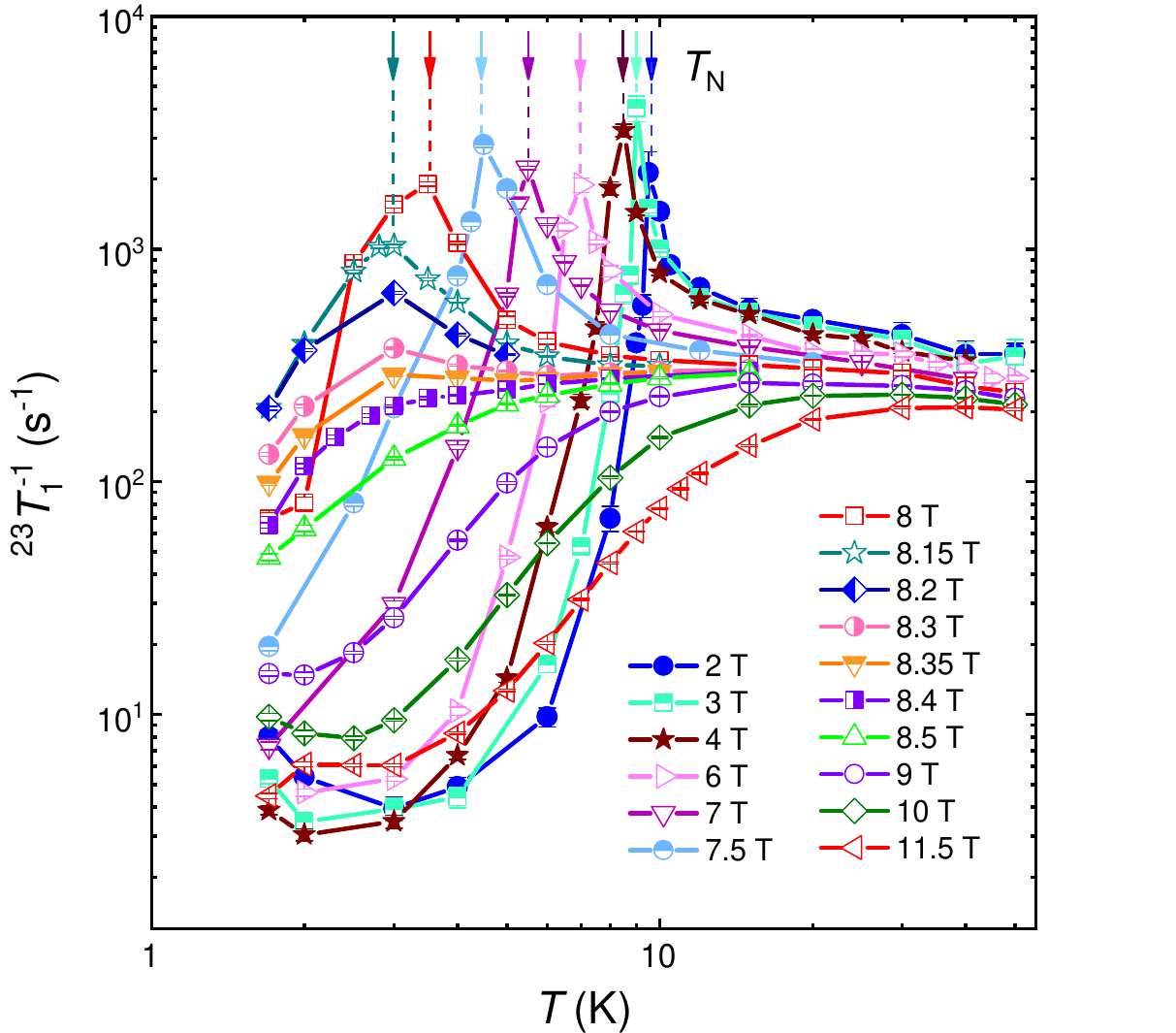}
\caption{\label{slrr}~\textbf{Spin-lattice relaxation rates.} $1/^{23}T_1$ as functions
of temperatures under different fields. The down
arrows mark the $T_N$ at the peak position in $1/^{23}T_1$.}
\end{figure}

\section{Spin-lattice relaxation rates and remnant low-energy spin fluctuations}

We further studied the spin dynamics by measuring the spin-lattice relaxation rates $1/^{23}T_1$.
Figure~\ref{slrr} demonstrates the $1/^{23}T_1$ as functions of temperatures
at different fields.
With fields from 2~T to 8~T, 1/$^{23}T_1$ increases when cooled from 50~K to 20~K, revealing strong AFM spin fluctuations due to intralayer
coupling at high temperatures. Upon further cooling, $1/^{23}T_1$ increases
sharply, which indicates the onset of 3D correlations. A peaked behavior
is established, marking the AFM phase transition.
From this, the $T_N$ are precisely determined and shown in the phase diagram in Fig.~\ref{pd}.
Below $T_N$, $1/^{23}T_1$ drops rapidly, demonstrating gapped spin wave excitations.
Such gapped behavior is also prominent with field above 9~T, when the  fully polarized phase
is achieved, as seen earlier in the magnetization data.
The critical field is determined to be $H_{C}$$\approx$~8.35~T, where the upturn
in $1/^{23}T_1$ just disappears.

However, for fields where the value of $1/^{23}T_1$ at 3~K is less than 20~s$^{-1}$,
either a level off or an upturn behavior is seen in $1/^{23}T_1$ with temperature below 3~K,
regardless of the ordered phase or the fully polarized phase. This behavior indicates
the existence of another type of low-energy fluctuations, being in contradiction to
intrinsic magnetic fluctuations where a fast drop of $1/^{23}T_1$ is expected.
To understand this, structural fluctuations may have to be considered. We speculate
that Na$^{+}$ movement plays an important role in causing the low-energy fluctuations.
In particular, Na$^{+}$ deficiency is known to exist in layered systems such as sodium
batteries, which may lead to low-energy structural fluctuations from Na$^{+}$ movement;
in addition, such structural fluctuations may disturb the magnetic moment and lead to
low-energy spin dynamics. To our knowledge, such behavior
has been rarely seen  in magnetic materials. We hope that further studies can clarify its nature and explore potential applications in sodium batteries.

\section{Spin gaps and Phase diagram}

\begin{figure}[t]
\includegraphics[width=8.5cm]{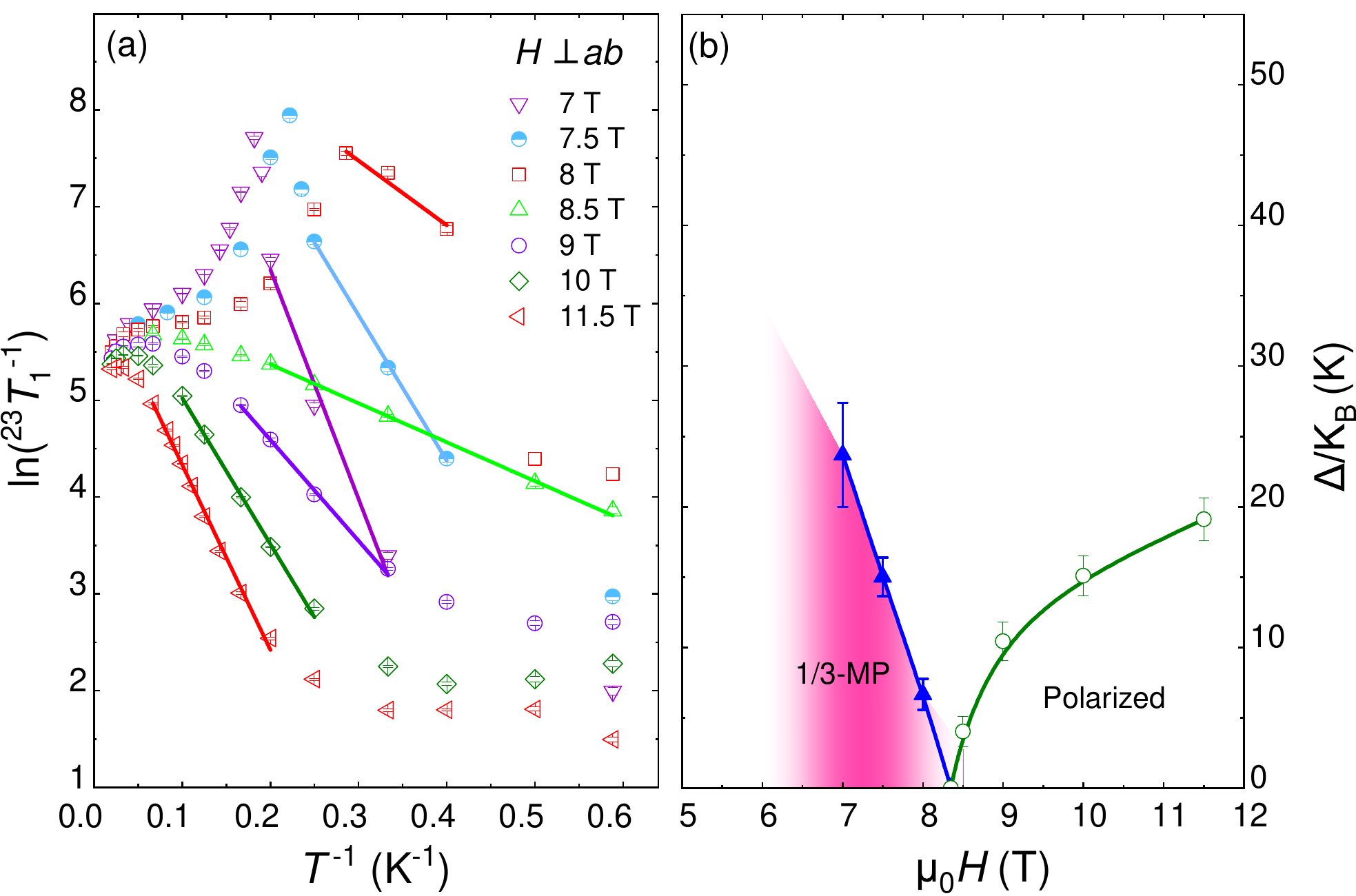}
\caption{\label{gap}~\textbf{Low-temperature spin gaps.} (a) ln(1/$^{23}T_1$) as functions of 1/$T$.
The straight lines are linear fit to the data in a limited temperature range to obtain the gaps ${\Delta}$ (see text).
(b) Spin gap at different fields.}
\end{figure}

\begin{figure}[h]
\includegraphics[width=8cm]{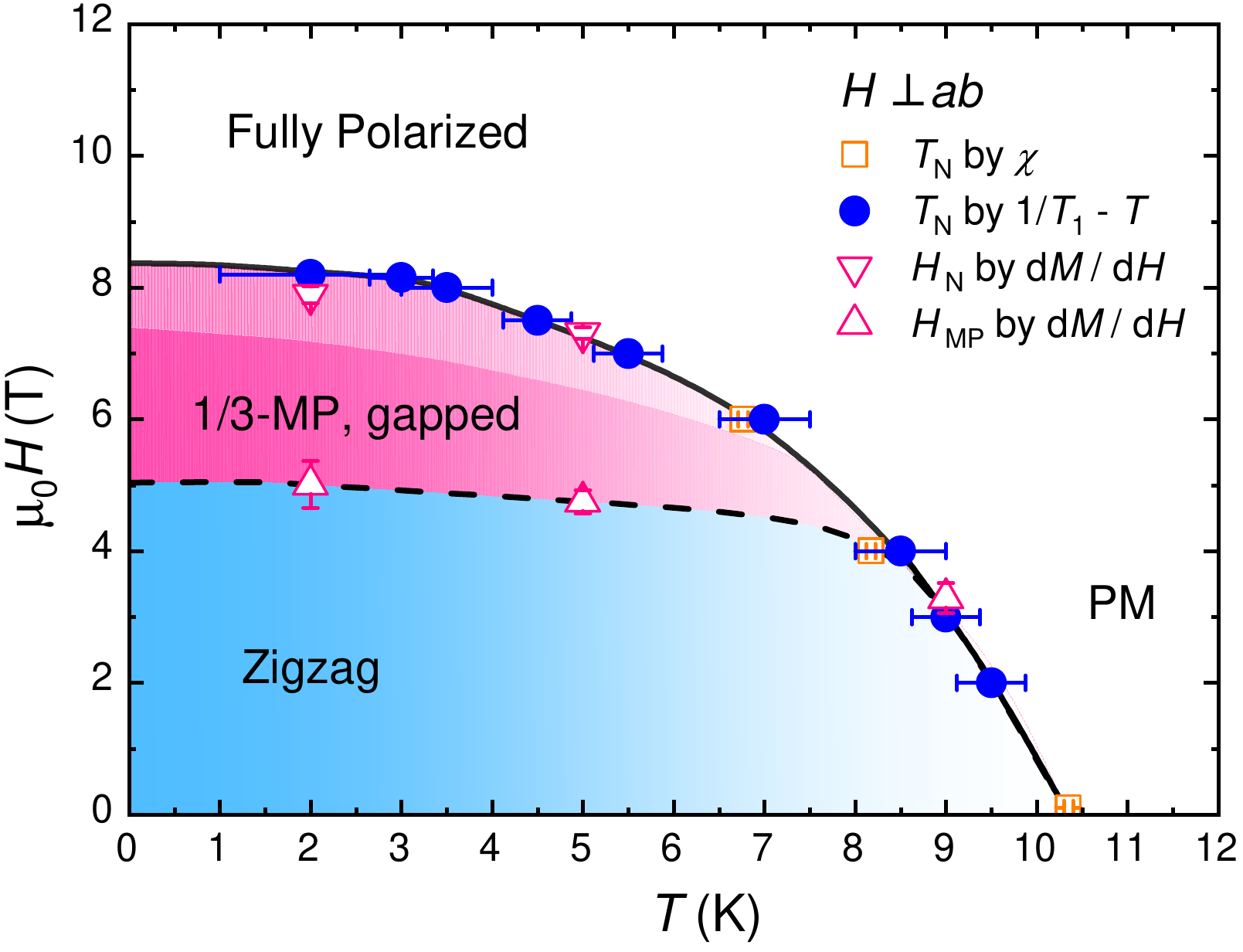}
\caption{\label{pd}~\textbf{Magnetic phase diagram with perpendicular field.}
The phase boundaries of the PM, the zigzag order, and the 1/3-MP are
determined with increasing field by different probes as labeled.
The dotted and the solid lines represent the first-order and the second-order phase transition, respectively.
}
\end{figure}

Now we study the gapped behavior in the ordered phase and the fully polarized phase.
In the following, we attempt to fit the low-temperature data
to the thermal activation function, that is, $1/T_1$$=$$ae^{-{\Delta}/k_BT}$,
where $\Delta$ is the spin gap.
To determine the gap values directly, ln($1/^{23}T_1$) is plotted against $1/T$
as shown in Fig.~\ref{gap}(a). In a limited regime,
a linear behavior is observed as shown by the straight line fitting, which gives the values of
$\Delta$  from the slope.  The obtained $\Delta$ are then plotted as a function of field
in Fig.~\ref{gap}(b). $\Delta$ first decreases with field in the 1/3-MP phase,
drops to zero at $H_{\rm C}$, and  increases again with field in the fully
polarized phase. Therefore, a quantum critical point is established at $H_{\rm C}$.
Note that because of the emergence of structural
fluctuations as discussed above, the fitting range for the gap is largely limited,
which may lead to overestimation on $\Delta$.

By summarizing aforementioned magnetic and NMR measurements results, a
complete ($H$,$T$) phase diagram is shown in Fig.~\ref{pd}.
A gapped, 1/3-MP phase is established,
separated from the zigzag order by a first-order phase transition,
and from the gapped, fully polarized phase by a second-order phase transition.

\section{Discussions and Summary}

The current work reveals several interesting observations in Na$_3$Ni$_2$BiO$_6$,
which include large, positive Curie-Weiss constant indicating existence
of strong ferromagnetic interactions, the zigzag order and a field-induced spin gapped, 1/3-MP phase
with both negative and large positive hyperfine fields on $^{23}$Na sites.
These results may help to resolve the exact magnetic structures of the 1/3-MP  phase.
Note that a fully polarized phase is established at a field
of 8.35~T, which is in contrast to other spin-1/2 Kitaev materials where a quantum
spin liquid may exist between the ordered phase and the fully polarized phase~\cite{Li_NC_2021}.

The existence of the 1/3-MP phase indicates strong exchange
frustration in the honeycomb plane. Combined with the observation of
ferromagnetic interactions, future studies  by inelastic
neutron scattering are highly
demanding to check if the proposed ferromagnetic Kitaev interactions exist~\cite{Wen_NP_2023}, and if other tuning methods are possible to access quantum spin liquid in this S=1
honeycomb lattice material.

In summary, we performed NMR study on a quasi-two-dimensional honeycomb lattice
antiferromagnet Na$_3$Ni$_2$BiO$_6$. Our NMR data demonstrate strong exchange frustration
by the existence of a 1/3-MP phase and ferromagnetic interactions,
which gives compelling possibilities for the presence of Kitaev interactions in a large spin system.
The spin gap, the first-order phase transition
to zigzag order, and the second-order phase transition to the fully polarized phase, are revealed in the plateau phase. In addition, we also observe strong low-energy fluctuations at
temperature below 3~K, which suggests a hidden degree of freedom yet to be understood.
These findings should provide significant understanding to the microscopic interactions and
competing quantum states in this material, and host possibility for achieving QSL in honeycomb
lattice with large spins.

{\bf Acknowledgments.} This work is supported by
the National Key R\&D Program of China (Grants Nos. 2023YFA1406500, 2022YFA1402700, and 2021YFA1400400) and the National Natural Science Foundation
of China (Grants Nos.~12134020, 12374156, 12104503, 12061131004, 12225407, and 12074174).

\bibliographystyle{iopart-num}
\bibliography{myref}

\providecommand{\newblock}{}
\begin{thebibliography}{10}
\expandafter\ifx\csname url\endcsname\relax
  \def\url#1{{\tt #1}}\fi
\expandafter\ifx\csname urlprefix\endcsname\relax\def\urlprefix{URL }\fi
\providecommand{\eprint}[2][]{\url{#2}}

\bibitem{Kitaev_AP_2006}
Kitaev A 2006 {\em Ann. Phys.\/} {\bf 321} 2

\bibitem{Khaliullin_PRL_2009}
Jackeli G and Khaliullin G 2009 {\em Phys. Rev. Lett.\/} {\bf 102} 017205

\bibitem{Kee_PRL_2014}
Rau J~G, Lee E~K~H and Kee H~Y 2014 {\em Phys. Rev. Lett.\/} {\bf 112} 077204

\bibitem{Kindo_PRB_2016}
Chanlert P, Kurita N, Tanaka H, Goto D, Matsuo A and Kindo K 2016 {\em Phys.
  Rev. B\/} {\bf 93} 094420

\bibitem{Hill_PRB_2011}
Liu X, Berlijn T, Yin W~G, Ku W, Tsvelik A, Kim Y~J, Gretarsson H, Singh Y,
  Gegenwart P and Hill J~P 2011 {\em Phys. Rev. B\/} {\bf 83} 220403

\bibitem{Cao_PRB_2012}
Ye F, Chi S, Cao H, Chakoumakos B~C, Fernandez-Baca J~A, Custelcean R, Qi T~F,
  Korneta O~B and Cao G 2012 {\em Phys. Rev. B\/} {\bf 85} 180403

\bibitem{Taylor_PRL_2012}
Choi S~K, Coldea R, Kolmogorov A~N, Lancaster T, Mazin I~I, Blundell S~J,
  Radaelli P~G, Singh Y, Gegenwart P, Choi K~R, Cheong S~W, Baker P~J, Stock C
  and Taylor J 2012 {\em Phys. Rev. Lett.\/} {\bf 108} 127204

\bibitem{Coldea_PRB_2015}
Johnson R~D, Williams S~C, Haghighirad A~A, Singleton J, Zapf V, Manuel P,
  Mazin I~I, Li Y, Jeschke H~O, Valent\'{\i} R and Coldea R 2015 {\em Phys.
  Rev. B\/} {\bf 92} 235119

\bibitem{Kim_PRB_2015}
Sears J~A, Songvilay M, Plumb K~W, Clancy J~P, Qiu Y, Zhao Y, Parshall D and
  Kim Y~J 2015 {\em Phys. Rev. B\/} {\bf 91} 144420

\bibitem{Simonet_PRB_2016}
Lefran\ifmmode~\mbox{\c{c}}\else \c{c}\fi{}ois E, Songvilay M, Robert J, Nataf
  G, Jordan E, Chaix L, Colin C~V, Lejay P, Hadj-Azzem A, Ballou R and Simonet
  V 2016 {\em Phys. Rev. B\/} {\bf 94} 214416

\bibitem{Ritter_PRB_2017}
Bera A~K, Yusuf S~M, Kumar A and Ritter C 2017 {\em Phys. Rev. B\/} {\bf 95}
  094424

\bibitem{Ling_JSSC_2016}
Wong C, Avdeev M and Ling C~D 2016 {\em J. Solid State Chem.\/} {\bf 243} 18

\bibitem{McGuire_PRM_2019}
Yan J~Q, Okamoto S, Wu Y, Zheng Q, Zhou H~D, Cao H~B and McGuire M~A 2019 {\em
  Phys. Rev. Mater.\/} {\bf 3} 074405

\bibitem{Park_JPCM_2022}
Kim C, Jeong J, Lin G, Park P, Masuda T, Asai S, Itoh S, Kim H~S, Zhou H, Ma J
  and Park J~G 2021 {\em J. Phys. Condens. Matter\/} {\bf 34} 045802

\bibitem{JiachengZheng_PRL_2017}
Zheng J, Ran K, Li T, Wang J, Wang P, Liu B, Liu Z~X, Normand B, Wen J and Yu W
  2017 {\em Phys. Rev. Lett.\/} {\bf 119} 227208

\bibitem{Nagler_NM_2016}
Li L, Stone M~B, Granroth G~E, Lumsden M~D, Yiu Y, Knolle J, Bhattacharjee S,
  Kovrizhin D~L, Moessner R, Tennant D~A, Mandrus D~G and Nagler S~E 2016 {\em
  Nat. Mater.\/} {\bf 15} 733

\bibitem{lee_PRL_2017}
Leahy I~A, Pocs C~A, Siegfried P~E, Graf D, Do S~H, Choi K~Y, Normand B and Lee
  M 2017 {\em Phys. Rev. Lett.\/} {\bf 118} 187203

\bibitem{Kim_PRB_2017}
Sears J~A, Zhao Y, Xu Z, Lynn J~W and Kim Y~J 2017 {\em Phys. Rev. B\/} {\bf
  95} 180411

\bibitem{Loidl_PRL_2017}
Wang Z, Reschke S, H\"uvonen D, Do S~H, Choi K~Y, Gensch M, Nagel U, R\~o\ om T
  and Loidl A 2017 {\em Phys. Rev. Lett.\/} {\bf 119} 227202

\bibitem{Nagler_PRB_2019}
Balz C, Lampen-Kelley P, Banerjee A, Yan J, Lu Z, Hu X, Yadav S~M, Takano Y,
  Liu Y, Tennant D~A, Lumsden M~D, Mandrus D and Nagler S~E 2019 {\em Phys.
  Rev. B\/} {\bf 100} 060405

\bibitem{Ong_NP_2021}
Czajka P, Gao T, Hirschberger M, Lampen-Kelley P, Banerjee A, Yan J, Mandrus
  D~G, Nagler S~E and Ong N~P 2021 {\em Nat. Phys.\/} {\bf 17} 915

\bibitem{Li_NC_2021}
Li H, Zhang H~K, Wang J, Wu H~Q, Gao Y, Qu D~W, Liu Z~X, Gong S~S and Li W 2021
  {\em Nat. Commun.\/} {\bf 12} 4007

\bibitem{Li_PRB_2020}
Yao W and Li Y 2020 {\em Phys. Rev. B\/} {\bf 101} 085120

\bibitem{Ma_NC_2021}
Lin G, Jeong J, Kim C, Wang Y, Huang Q, Masuda T, Asai S, Itoh S, G¨¹nther G,
  Russina M, Lu Z, Sheng J, Wang L, Wang J, Wang G, Ren Q, Xi C, Tong W, Ling
  L, Liu Z, Wu L, Mei J, Qu Z, Zhou H, Wang X, Park J~G, Wan Y and Ma J 2021
  {\em Nat. Commun.\/} {\bf 12} 5559

\bibitem{Li_PRX_2022}
Li X, Gu Y, Chen Y, Garlea V~O, Iida K, Kamazawa K, Li Y, Deng G, Xiao Q, Zheng
  X, Ye Z, Peng Y, Zaliznyak I~A, Tranquada J~M and Li Y 2022 {\em Phys. Rev.
  X\/} {\bf 12} 041024

\bibitem{ZeHu_PRB_2024}
Hu Z, Chen Y, Cui Y, Li S, Li C, Xu X, Chen Y, Li X, Gu Y, Yu R, Zhou R, Li Y
  and Yu W 2024 {\em Phys. Rev. B\/} {\bf 109} 054411

\bibitem{Wen_NP_2023}
Shangguan Y, Bao S, Dong Z~Y, Xi N, Gao Y~P, Ma Z, Wang W, Qi Z, Zhang S, Huang
  Z, Liao J, Zhao X, Zhang B, Cheng S, Xu H, Yu D, Mole R~A, Murai N,
  Ohira-Kawamura S, He L, Hao J, Yan Q~B, Song F, Li W, Yu S~L, Li J~X and Wen
  J 2023 {\em Nat. Phys.\/} {\bf 19} 1883

\bibitem{Winkelmann_JPCM_1994}
Schotte U, Stusser N, Schotte K~D, Weinfurter H, Mayer H~M and Winkelmann M
  1994 {\em J. Phys. Condens. Matter\/} {\bf 6} 10105

\bibitem{Goto_PRB_2003}
Ono T, Tanaka H, Aruga~Katori H, Ishikawa F, Mitamura H and Goto T 2003 {\em
  Phys. Rev. B\/} {\bf 67} 104431

\bibitem{Takano_PRB_2007}
Tsujii H, Rotundu C~R, Ono T, Tanaka H, Andraka B, Ingersent K and Takano Y
  2007 {\em Phys. Rev. B\/} {\bf 76} 060406

\bibitem{Takano_PRL_2009}
Fortune N~A, Hannahs S~T, Yoshida Y, Sherline T~E, Ono T, Tanaka H and Takano Y
  2009 {\em Phys. Rev. Lett.\/} {\bf 102} 257201

\bibitem{Kindo_PRL_2012}
Shirata Y, Tanaka H, Matsuo A and Kindo K 2012 {\em Phys. Rev. Lett.\/} {\bf
  108} 057205

\bibitem{Takano_PRL_2012}
Zhou H~D, Xu C, Hallas A~M, Silverstein H~J, Wiebe C~R, Umegaki I, Yan J~Q,
  Murphy T~P, Park J~H, Qiu Y, Copley J~R~D, Gardner J~S and Takano Y 2012 {\em
  Phys. Rev. Lett.\/} {\bf 109} 267206

\bibitem{Tanaka_PRL_2013}
Susuki T, Kurita N, Tanaka T, Nojiri H, Matsuo A, Kindo K and Tanaka H 2013
  {\em Phys. Rev. Lett.\/} {\bf 110} 267201

\bibitem{Ma_NC_2018}
Kamiya Y, Ge L, Hong T, Qiu Y, Quintero-Castro D~L, Lu Z, Cao H~B, Matsuda M,
  Choi E~S, Batista C~D, Mourigal M, Zhou H~D and Ma J 2018 {\em Nat.
  Commun.\/} {\bf 9} 2666

\bibitem{Zhitomirsky_PRL_2002}
Zhitomirsky M~E 2002 {\em Phys. Rev. Lett.\/} {\bf 88} 057204

\bibitem{Senthil_PRL_2006}
Damle K and Senthil T 2006 {\em Phys. Rev. Lett.\/} {\bf 97} 067202

\bibitem{Hotta_NC_2013}
Nishimoto S, Shibata N and Hotta C 2013 {\em Nat. Commun.\/} {\bf 4} 2287

\bibitem{Kumar_PRB_2021}
Adhikary M, Ralko A and Kumar B 2021 {\em Phys. Rev. B\/} {\bf 104} 094416

\bibitem{Suwa_PRB_2022}
Gen M and Suwa H 2022 {\em Phys. Rev. B\/} {\bf 105} 174424

\bibitem{Notych_SSC_1993}
Lozovik Y and Notych O 1993 {\em Solid State Commun.\/} {\bf 85} 873

\bibitem{Ueda_PRL_1999}
Kageyama H, Yoshimura K, Stern R, Mushnikov N~V, Onizuka K, Kato M, Kosuge K,
  Slichter C~P, Goto T and Ueda Y 1999 {\em Phys. Rev. Lett.\/} {\bf 82} 3168

\bibitem{Mila_Sci_2002}
Kodama K, Takigawa M, Horvati\'{c} M, Berthier C, Kageyama H, Ueda Y, Miyahara
  S, Becca F and Mila F 2002 {\em Science\/} {\bf 298} 395

\bibitem{Golosov_JPCM_1991}
Chubukov A~V and Golosov D~I 1991 {\em J. Phys. Condens. Matter\/} {\bf 3} 69

\bibitem{Honecker_JPCM_1999}
Honecker A 1999 {\em J. Phys. Condens. Matter\/} {\bf 11} 4697

\bibitem{Starykh_RPP_2015}
Starykh O~A 2015 {\em Rep. Prog. Phys.\/} {\bf 78} 052502

\bibitem{Petrenko_PRL_2000}
Zhitomirsky M~E, Honecker A and Petrenko O 2000 {\em Phys. Rev. Lett.\/} {\bf
  85} 3269

\bibitem{Miyashita_JPSJ_1985}
Kawamura H and Miyashita S 1985 {\em J. Phys. Soc. Jpn\/} {\bf 54} 4530

\bibitem{Henley_PRL_1989}
Henley C~L 1989 {\em Phys. Rev. Lett.\/} {\bf 62} 2056

\bibitem{Starykh_PRL_2009}
Alicea J, Chubukov A~V and Starykh O~A 2009 {\em Phys. Rev. Lett.\/} {\bf 102}
  137201

\bibitem{Mila_PRB_2013}
Coletta T, Zhitomirsky M~E and Mila F 2013 {\em Phys. Rev. B\/} {\bf 87} 060407

\bibitem{Danshita_PRL_2014}
Yamamoto D, Marmorini G and Danshita I 2014 {\em Phys. Rev. Lett.\/} {\bf 112}
  127203

\bibitem{Goto_JPSJ_1996}
Inami T, Ajiro Y and Goto T 1996 {\em J. Phys. Soc. Jpn\/} {\bf 65} 2374

\bibitem{Nakano_JPSJ_2011}
Shirata Y, Tanaka H, Ono T, Matsuo A, Kindo K and Nakano H 2011 {\em J. Phys.
  Soc. Jpn\/} {\bf 80} 093702

\bibitem{Schlottmann_PRL_2012}
Hwang J, Choi E~S, Ye F, Dela~Cruz C~R, Xin Y, Zhou H~D and Schlottmann P 2012
  {\em Phys. Rev. Lett.\/} {\bf 109} 257205

\bibitem{Shannon_PRB_2011}
Seabra L, Momoi T, Sindzingre P and Shannon N 2011 {\em Phys. Rev. B\/} {\bf
  84} 214418

\bibitem{Hagiwara_JPSJ_2019}
Okutani A, Kida T, Narumi Y, Shimokawa T, Honda Z, Kindo K, Nakano T, Nozue Y
  and Hagiwara M 2019 {\em J. Phys. Soc. Jpn\/} {\bf 88} 013703

\bibitem{Cava_Inorg_2013}
Seibel E~M, Roudebush J~H, Wu H, Huang Q, Ali M~N, Ji H and Cava R~J 2013 {\em
  Inorg. Chem.\/} {\bf 52} 13605

\end{thebibliography}

\end{document}